\documentclass{article}
\usepackage{ijcai26}

\usepackage{times}
\usepackage{soul}
\usepackage{url}
\usepackage[hidelinks]{hyperref}
\usepackage[utf8]{inputenc}
\usepackage[small]{caption}
\usepackage{capt-of}
\usepackage{graphicx}
\usepackage{amsmath}
\usepackage{amssymb}
\usepackage{amsthm}
\usepackage{booktabs}
\usepackage{color}
\usepackage[switch]{lineno}
\usepackage[linesnumbered,ruled]{algorithm2e}
\usepackage{silence}
\usepackage{microtype}
\theoremstyle{plain}

\newtheorem{definition}{Definition}

\newcommand{\nauty}{\textsc{nauty}}

\makeatletter
\g@addto@macro\bfseries{\boldmath}
\renewcommand{\SetKwInOut}[2]{%
  \sbox\algocf@inoutbox{\KwSty{#2\algocf@typo}}
  \expandafter\ifx\csname InOutSizeDefined\endcsname\relax
    \newcommand\InOutSizeDefined{}\setlength{\inoutsize}{\wd\algocf@inoutbox}%
    \sbox\algocf@inoutbox{\parbox[t]{\inoutsize}{\KwSty{#2\algocf@typo\hfill}}~}\setlength{\inoutindent}{\wd\algocf@inoutbox}
  \else
    \ifdim\wd\algocf@inoutbox>\inoutsize%
    \setlength{\inoutsize}{\wd\algocf@inoutbox}%
    \sbox\algocf@inoutbox{\parbox[t]{\inoutsize}{\KwSty{#2\algocf@typo\hfill}}~}\setlength{\inoutindent}{\wd\algocf@inoutbox}
    \fi%
  \fi%
  \algocf@newcommand{#1}[1]{%
    \ifthenelse{\boolean{algocf@hanginginout}}{\relax}{\algocf@seteveryparhanging{\relax}}%
    \ifthenelse{\boolean{algocf@inoutnumbered}}{\relax}{\algocf@seteveryparnl{\relax}}%
    {\let\\\algocf@newinout\hangindent=\inoutindent\hangafter=1\parbox[t]{\inoutsize}{\KwSty{#2\algocf@typo\hfill}}~##1\par}
    \algocf@linesnumbered
  }}%
\makeatother

\title{SAT + {\nauty}: Orderly Generation of Small Kochen-Specker Sets Containing the Smallest State-independent Contextuality Set}

\author{
Zhengyu Li$^1$\and
Curtis Bright$^2$\and
Stefan Trandafir$^3$\and
Ad\'an~Cabello$^{3,4}$\And
Vijay Ganesh$^{1}$\\
\affiliations
$^1$Georgia Institute of Technology, Atlanta, USA\\
$^2$University of Waterloo, Canada\\
$^3$Departamento de F\'{\i}sica Aplicada II, Universidad de Sevilla, E-41012 Sevilla, Spain\\
$^4$Instituto Carlos~I de F\'{\i}sica Te\'orica y Computacional, Universidad de
Sevilla, E-41012 Sevilla, Spain\\
\emails
brian.li@gatech.edu,
cbright@uwaterloo.ca,
strandafir@us.es,
adan@us.es
vganesh45@gatech.edu,
}

\begin{document}

\maketitle

\begin{abstract}
    We present a search for small Kochen-Specker (KS) sets in dimension $3$, specifically targeting extensions of the $13$-ray Yu-Oh set, which has been proven to be the minimal witness to state-independent contextuality.
    To enable this search, we introduce a novel SAT-based orderly generation framework integrating recursive canonical labeling (RCL) with the graph isomorphism tool {\nauty}.
    We demonstrate that previous SAT approaches relying on lexicographical canonicity suffer from exponential scaling on canonical graphs.
    This limitation renders them intractable on the large instances (25 to 33 vertices) encountered in our search, whereas our RCL check maintains consistent millisecond-level performance ($0.005$s), effectively eliminating the bottleneck.
    Overcoming this bottleneck allows us to perform the first exhaustive enumeration of all KS sets with up to $33$ rays containing the complete $25$-ray state-independent contextuality (SI-C) set obtained by rigid extensions of the Yu-Oh set in $1{,}641$ CPU hours.
    We found and verified the $33$-ray set discovered by Sch\"utte is the smallest three-dimensional KS set containing the complete 25-ray SI-C set.
    All non-existence results are backed by independently verifiable proof certificates via an extension of the DRAT proof format. 
\end{abstract}

\section{Introduction}

A Kochen-Specker (KS) set is a finite set of rays in $\mathbb{C}^d$ for some $d \ge 3$, which does not admit a $\{0,1\}$-assignment of the rays such that (i) each orthogonal basis contains exactly one ray assigned $1$, and (ii) in each pair of orthogonal rays, at most one ray is assigned $1$.

A state-independent contextuality (SI-C) set is a set of rays $\mathbb{C}^d$ for some $d \ge 3$ for which the corresponding rank-one observables have some non-contextual inequality that is violated by any initial quantum state. 

Here we restrict our attention to the smallest possible dimension, $3$. 
It has been proven \cite{Cabello:2016JPA} that the smallest SI-C set contains $13$ rays in dimension~$3$ and consists of all real vectors with coordinates in $\{0,1,-1\}$~\cite{Yu:2012PRL}. In contrast, it is not known what the smallest KS set is in dimension~$3$: a problem which has been open for just under 60 years since the discovery of the first such set by Kochen and Specker \cite{Kochen:1965a}. The smallest known is due to Conway and Kochen and has $31$ rays. Exhaustive searches have proven that the smallest KS set in dimension 3 must have at least $24$ rays (see \citeauthor{Kirchweger:2023} and \citeauthor{Li:2024}).

All KS sets are SI-C sets, but in general KS sets may contain small SI-C sets that are not KS sets. Indeed, each of the small known Kochen-Specker sets in dimension 3 contains a copy of the $13$-ray SI-C set. Moreover, the other small 3-dimensional SI-C set in the literature has $21$ rays \cite{Bengtsson:2012PLA}, and the smallest known KS set containing it has $55$ rays \cite{Trandafir:2025PRAa}. 

Recently, using the $13$-ray Yu-Oh set as a starting point, a new Kochen-Specker set has been discovered that also has $33$ rays, but has only $14$ orthogonal bases~\cite{Cabello:2025XXX}. Given that the history of Kochen-Specker sets is nearly sixty years old~\cite{Kochen:1967JMM}, it is surprising that such a fundamental object remained undiscovered until recently.

It has recently been shown that Kochen-Specker sets correspond exactly to bipartite nonlocal games for which there exists a perfect quantum strategy (BPQS) \cite{Cabello:2025PRL}. These are games played between two players, Alice and Bob, who have input sets $X$ and $Y$ (respectively), and output sets $A$ and $B$ (respectively), and for which each choice of inputs and corresponding outputs is either winning or losing. Kochen-Specker sets provide games and accompanying optimal quantum strategies for Alice and Bob, allowing them to win at each round of the game, while no such classical strategy exists \cite{CinelliPRL2005,YangPRL2005,Aolita:2012PRA,Xu:2022PRL,kumar2025}.
Small Kochen-Specker sets play a very important role in this setting since they lead to games with low input cardinality $|X||Y|$ (and thus more experimentally feasible scenarios). For example, the aforementioned $33$ ray set produces the smallest known input cardinality for dimension 3.

Recently a particular type of SI-C set (or KS set), called a \emph{rigid} SI-C (KS) set has proven to be especially important (in essence the orthogonalities of such a set are unique up to unitary transformations).
\citeauthor{Xu:2024PRL} showed that rigid SI-C sets enable certification with any full-rank state (CFR), which simplify preparation of quantum experiments and improve robustness under experimental imperfections. Moreover, such KS sets provide the only known way to self-test supersinglets of $d$ particles at $d$ levels. Naturally, there has been much interest in identifying small rigid KS sets (see for example \cite{Aravind:2025,Trandafir:2025PRAa,Kernaghan:2026}). 



We focus on the construction method where, starting with the Yu-Oh 13-ray SI-C set, we recursively choose pairs of rays and add the unique ray orthogonal to the pair. One may construct the $31$-ray and $37$-ray~\cite{Peres:1993} sets of Conway and Kochen in this manner, as well as the $33$-ray set of Schütte \cite{Bub:1996FP,Peres:1993}. The aplication of this procedure to every pair of the Yu-Oh set yields a rigid set containing $25$ rays. This $25$-ray core is the natural place to implement a search for the smallest rigid KS set in dimension 3, since such a search will necessarily find any rigid KS set containing the smallest SI-C set.

\begin{figure}[!t]
    \centering
    \includegraphics[width=0.92\linewidth]{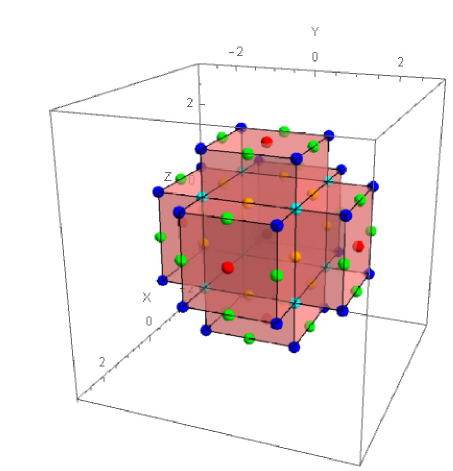}
    \caption{If we connect the center of the figure with all the dots, we get $37$ directions (rays) whose components are $(0,0,1)$, $(0,1,\pm1)$, $(0,1,\pm2)$, $(1,1,\pm1)$, $(1,\pm1,\pm2)$ and their permutations. Schütte-33 is obtained by removing the rays $(0,1,\pm2)$ and $(0,2,\pm1)$.}
    \label{fig:Schutte-33}
\end{figure}

Our goal in this work is to apply exhaustive search in order to find (rigid) Kochen-Specker sets in dimension~$3$ obtained from this SI-C set.  The main contributions of our work are:
\begin{itemize}
    \item \textbf{A new canonical form that leverages the empirical efficiency of {\nauty}~\cite{MCKAY201494}, a tool\footnote{Open source software available at \url{https://github.com/BrianLi009/SAT-nauty}.} to compute graph automorphism groups, while preserving the hierarchical property required for SAT-based orderly generation.} 
    Previous SAT-based approaches to orderly generation rely on lexicographical definitions of canonicity, which fail to scale to graphs of large order $n$. Conversely, while specialized tools like {\nauty} are highly efficient at isomorphism checking, they cannot be directly embedded into a SAT solver to perform orderly generation. This is because the standard canonical labeling computed by {\nauty} is not \emph{hereditary} (or prefix-preserving): a canonical graph may possess non-canonical subgraphs. Consequently, a solver cannot simply use {\nauty} to prune partial assignments, as this would discard valid branches that eventually lead to canonical solutions. We combine the best of both worlds by implementing Recursive Canonical Labeling (RCL), a technique introduced by Afzaly~\cite{afzaly2016generation} that constructs a hierarchical canonical form from any base labeling function. By building RCL on top of {\nauty}, we create a \textbf{SAT+\nauty} framework that successfully utilizes {\nauty}'s empirical efficiency within a valid orderly generation scheme.
    A SAT + graph isomorphism tool was previously used in the resolution of Lam's problem~\cite{Bright2021} via recording an isomorphism certificate for every graph explored. However, the list of recorded objects can grow large, and to effectively exploit parallelization it needs to be shared across all processors, a limitation that is not present in our approach.

    \item \textbf{Exhaustive enumeration of KS sets extending the complete SI-C set.} 
    We enumerate all Kochen-Specker sets up to $33$ rays containing the complete SI-C set of order $25$. This set is the unique closure of the original $13$-ray Yu-Oh set under the operation of adding orthogonal rays to existing pairs. While general recursive extension from the $13$-ray set yields a vast search space, we focus on this dense $25$-ray core. Our search confirms that, up to order $33$, the $33$-ray KS set by Schütte~\cite{Bub:1996FP,Peres:1993} is the \emph{only} KS set containing this complete $25$-ray structure. The set is illustrated in Figure \ref{fig:Schutte-33} as a subset of the $37$-ray set of Conway and Kochen (which was unpublished, but appears in \cite{Cabello:1996}). This classification is fundamental to quantum foundations, as the Yu-Oh set constitutes the minimal witness to state-independent contextuality. By exhaustively mapping its rigid extensions, we rigorously define the geometric boundaries of the simplest contextual structures in dimension three.
\end{itemize}

\section{Related Work}

The traditional approach to prevent a SAT solver from repeatedly exploring isomorphic parts of a search space is via the use of \emph{symmetry breaking} techniques.
One such symmetry breaking approach is to add ``static'' constraints prior to the search that reduces the size of the search space~\cite{crawford1996symmetry,Heule2019}.  Another approach is to ``dynamically''
break symmetries during the search~\cite{DBLP:conf/ijcai/SellmannH05,Metin2018}.
Within the SAT context, two prominent frameworks that use dynamic symmetry breaking in this domain are the SAT+CAS paradigm~\cite{bright2022satisfiability} and the SAT Modulo Symmetries (SMS) framework~\cite{KirchwegerS24}.

\paragraph{SAT+CAS.}
\citeauthor{Li:2024}\ applied the SAT+CAS paradigm to the Kochen-Specker problem, establishing a verified lower bound of $24$ for the size of a KS system. This approach combines a SAT solver (MapleSAT~\cite{liang2016learning}) with a Computer Algebra System (CAS) to perform isomorph-free orderly generation. In this framework, canonicity is defined lexicographically based on the concatenation of the above-diagonal entries of the adjacency matrix columns. The solver prunes the search space by blocking partial assignments that are detected to be non-canonical.

\paragraph{SAT Modulo Symmetries (SMS).}
\citeauthor{Kirchweger:2023}\ independently established the lower bound of $24$ using the SMS framework. The tool integrates symmetry breaking into the CDCL loop of the SAT solver via a custom propagator. Unlike in \citeauthor{Li:2024}'s SAT+CAS approach, SMS defines canonicity based on the lexicographical minimality of the flattened row-wise adjacency matrix. The propagator performs a minimality check on partial assignments to ensure they can be extended to a canonical graph.

\paragraph{Limits of Lex-based Canonicity.}
Both SAT+CAS and SMS rely on the lexicographical definitions of canonicity. Consequently, neither framework addresses the intrinsic scalability bottleneck associated with these checks. To strictly guarantee that a given graph $G$ is canonical under a lexicographical definition, one must essentially verify that no other permutation of vertices yields a lexicographically smaller adjacency string. In the worst case, this requires testing against the space of all $n!$ permutations, which becomes computationally infeasible as $n$ grows. This shared limitation underscores the advantage of the Recursive Canonical Labeling (RCL) approach, which circumvents the need for such exhaustive permutation checks by leveraging the efficiency of {\nauty} while maintaining the hereditary property required for effective pruning. 

\section{Definitions}\label{sec:definitions}

A \emph{state-independent contextuality} (SI-C) set in dimension $d$ is a set of rank-one ideal observables (or, equivalently rays) violating a non-contextual inequality for any initial state in dimension $d$.

A {\em Kochen-Specker (KS) set} \cite{Kochen:1967JMM} is a finite set of rank-one observables $\mathcal{V}$ (or, equivalently, rays) in a Hilbert space ${\cal H}=\mathbb{C}^d$ of finite dimension $d \ge 3$, which does not admit an assignment $f\colon \mathcal{V} \rightarrow \{0,1\}$ such that $f(u) + f(v) \leq 1$ for $u, v \in \mathcal{V}$ orthogonal, and $\sum_{u \in b} f(u) = 1$ for every orthonormal basis $b \subseteq \mathcal{V}$.

Given a set $K$ of vectors in $\mathbb{C}^d$, the \emph{orthogonality graph} of $K$ is the graph whose vertices are the vectors of $K$ and for which two vertices are adjacent if and only if their corresponding vectors are orthogonal.

A graph $G$ is \emph{$010$-colorable} if there is a $\{0,1\}$-coloring of the vertices such that the following two conditions are satisfied simultaneously:

a. Adjacent vertices are not both colored $1$.
b. For each triangle in $G$, there is exactly one vertex that is colored $1$.

A set of vectors does not form a Kochen-Specker set if its orthogonality graph is $010$-colorable. On the other hand, given a graph that is non-$010$-colorable, it is not immediate that there is an associated Kochen-Specker set. This is only the case if there is a suitable set of vectors whose orthogonalities are captured by that graph (this is called a \emph{faithful orthogonal representation} of the graph $G$; for brevity, we refer to a faithful orthogonal representation simply as an orthogonal representation). 

\section{Methodology}\label{sec:methodology}

In this work, we search for Kochen-Specker (KS) sets in dimension $3$ with at most $33$ rays containing the Yu-Oh $13$-ray SI-C set. We focus on a specific type of subgraph constructions relevant to quantum foundations, followed by a SAT-based encoding to exhaustively enumerate valid KS sets.

\subsection{Recursive Extension from the Complete SI-C Set.}

The approach builds upon the observation that the $13$-ray Yu-Oh set can be naturally extended by adding the unique ray orthogonal to any pair of existing non-parallel rays.
The ray orthogonal to two rays is given by the cross product of those two rays.
Though the cross product vector itself is scale-dependent, the ray it spans is uniquely defined because we care only about the direction
of the vector, not its length. 
Repeatedly applying this operation to closure yields a unique set of $25$ rays, which we term the complete SI-C set.
Famous examples of KS sets, such as the $33$-ray set of Schütte, are known to contain this specific $25$-ray substructure.
In this work, we specifically target this class of systems. We fix the $25$-ray complete set as a base subgraph and employ our SAT solver to exhaustively enumerate all Kochen-Specker extensions up to order $33$. This narrows the search space to the most structured candidates while ensuring we find any variant of the Schütte set that might exist.

\subsection{SAT Encoding}\label{sec:encoding}
To perform the search, we encode the properties of a KS graph into a Boolean Satisfiability (SAT) instance. We employ the encoding detailed in the previous SAT+CAS framework~\cite{Li:2024}, which formulates the encoding via Boolean variables $e_{i,j}$ (representing vertices $i$ and $j$ are adjacent, i.e., the rays $i$ and $j$ are orthogonal) and $t_{i,j,k}$ (representing rays $i$, $j$, $k$ form a mutually orthogonal triple).

We enforce the following constraints to prune the search space effectively:
\begin{enumerate}
    \item \textbf{Structural Constraints:} We leverage properties from graph theory that a minimal KS set must satisfy, specifically:
    \begin{itemize}
        \item \textbf{Squarefree:} The graph must not contain any 4-cycles ($C_4$).
        \item \textbf{Minimum Degree:} Every vertex must have a degree of at least $3$.
        \item \textbf{Triangle Property:} Every vertex must be part of at least one triangle (a mutually orthogonal triple).
    \end{itemize}
    \item \textbf{Non-010-Colorability:} The core KS property requires that the graph admits no 010-coloring. This is encoded by generating blocking clauses for all valid 010-colorings. Following the optimization in~\cite{Li:2024}, we only generate clauses for colorings where the set of vertices $V_1$ assigned the value $1$ is small (specifically, $|V_1| < \lceil n/2 \rceil$). We prioritize blocking these cases because colorings with a large number of $1$s (large $|V_1|$) are unlikely to be 010-colorable, rendering explicit blocking clauses for them unnecessary.
    \item \textbf{Fixed Subgraphs:} We first compute the canonical labeling of the starting configuration (of order $n = 25$) and fix the corresponding edge variables via unit clauses. This initialization step highlights a critical scalability advantage over lexicographical approaches (like SAT+CAS): merely computing the canonical form of these large base subgraphs is difficult under a pure lexicographical definition, which would prevent such solvers from effectively initializing the search space.
\end{enumerate}

Any satisfying assignment to this formula corresponds to a valid KS candidate graph. The geometric realizability of these candidates is then verified using the orthogonality check described in Section~\ref{sec:orthogonality}.

\section{SAT + \nauty-based Orderly Generation}
\label{sec:sat-nauty}

A key component of our pipeline is the integration of a SAT solver with an isomorph-free orderly generation routine. The orderly generation approach (developed in 1978 in two independent publications~\cite{read1978every,faradvzev1978constructive}) allows for the exhaustive enumeration of non-isomorphic graphs without the need to store previously generated isomorphism classes. This technique is the foundation for efficient combinatorial generators such as \textsc{genreg}~\cite{meringer1999fast}, which prune the search tree by enforcing that every intermediate graph must be the canonical representative of its isomorphism class.

Classical orderly generation relies on a specific definition of canonicity that satisfies the \emph{hereditary property}: if a graph is canonical, then its parent (typically the induced subgraph obtained by removing the last vertex) must also be canonical. This property implies the following:
\begin{quote}
    \textit{If a partial graph (prefix) is non-canonical, no extension of that graph can ever become canonical.}
\end{quote}
This allows the search to prune non-canonical intermediates immediately. However, highly efficient canonical labeling tools like {\nauty}~\cite{mckay2007nauty} produce labelings that are not hereditary, making them unsuitable for direct use in orderly generation.

To resolve this, we implement \emph{Recursive Canonical Labeling} (RCL), a technique introduced by Afzaly~\cite{afzaly2016generation}. RCL acts as a wrapper that transforms \emph{any} base canonical labeling function $\delta$ into a hierarchical (hereditary) canonical labeling. In our implementation, we use {\nauty} as the base function $\delta$ due to its efficiency, but the framework is agnostic to this choice.

\subsection{Recursive Canonical Labeling}

\begin{algorithm}[t!]
\caption{SAT + {\nauty} Orderly Generation. The \textsc{Solve} routine initializes the solver and registers the symmetry check. \textsc{CheckCanonicity} takes in partial assignments to verify that the induced subgraph $G_k$ is canonical. If $G_k$ is not canonical, a blocking clause is added to the solver immediately, forcing it to backtrack. \textsc{RecursiveCanonical} recursively constructs the hereditary canonical form.}\label{alg:sat-nauty}
\DontPrintSemicolon

\SetKwSty{textsc}
\SetKwInOut{Input}{Input}
\SetKwInOut{Output}{Output}

\SetKwFunction{Solve}{Solve}
\SetKwFunction{CheckCanonicity}{CheckCanonicity} 
\SetKwFunction{RecursiveCanonical}{RecursiveCanonical}
\SetKwFunction{GetCompletePrefix}{GetCompletePrefix}
\SetKwFunction{Nauty}{Nauty}
\SetKwFunction{AddBlockingClause}{AddBlockingClause}

\Input{$n$: number of vertices; $\Phi$: problem constraints (CNF)}
\Output{All canonical graphs satisfying $\Phi$}

\SetKwProg{Fn}{Function}{:}{}
\Fn{\Solve{$\Phi$, $n$}}{
    Initialize SAT solver with formula $\Phi$\;
    Register \CheckCanonicity as callback\;
    \While{SAT solver finds satisfying assignment $\sigma$}{
        Output $G_\sigma$\;
        Add blocking clause $\neg \sigma$\;
    }
}

\BlankLine
\Fn{\CheckCanonicity{partial assignment $\pi$}}{
    $k \gets$ \GetCompletePrefix{$\pi$}\;
    \If{$k \geq 2$ and $k$ increased since last check}{
        $G_k \gets$ subgraph induced by $\{1, \ldots, k\}$ under $\pi$\;
        $(G_k^{\text{can}}, \gamma) \gets$ \RecursiveCanonical{$G_k$}\;
        
        \If{$G_k^{\text{can}} \neq G_k$}{
            $C \gets$ clause blocking assignment of $G_k$\;
            \AddBlockingClause{$C$, witness $\gamma$}\;
        }
    }
}

\BlankLine
\Fn{\RecursiveCanonical{$G_k$}}{
    \If{$k = 1$}{
        \Return $(G_k, \text{identity})$\;
    }
    $\delta \gets$ \Nauty{$G_k$}\;
    $v \gets$ vertex with $\delta(v) = k$\;
    $G_{k-1} \gets G_k[V(G_k) \setminus \{v\}]$\;
    $(G_{k-1}^{\text{can}}, \gamma') \gets$ \RecursiveCanonical{$G_{k-1}$}\;
    Extend $\gamma'$ to $\gamma$ by setting $\gamma(v) = k$\;
    \Return $(G_k^{\gamma}, \gamma)$\;
}
\end{algorithm}

We first formally define the hierarchical property required for our search, and then describe the Recursive Canonical Labeling (RCL) algorithm used to satisfy it. Let $G$ be a graph on $n$ vertices labeled $\{1, \dots, n\}$. We denote $G[k]$ as the induced subgraph of $G$ on the vertices $\{1, \dotsc, k\}$.

\begin{definition}[Hierarchical Canonical Labeling]
A canonical labeling function $C$ is \emph{hierarchical} if for any graph $G$ that is canonical under $C$ (i.e., $C(G) = G$), and for any $1 \le k < n$, the prefix $G[k]$ is also canonical under $C$.
\end{definition}

RCL constructs such a labeling recursively. We start with an arbitrary \emph{base canonizer} $\delta$---defined as any function that computes a canonical labeling for a graph (i.e., ensuring that if $G \cong H$, their labeled forms are identical), regardless of whether it satisfies the hierarchical property. In our implementation, we use {\nauty} as this base canonizer due to its efficiency.

Given a graph $G$ and the base canonizer $\delta$, the Recursive Canonical Labeling is computed as follows:
\begin{enumerate}
    \item Compute the base labeling $\pi = \delta(G)$ using the underlying tool.
    \item Identify the vertex $v$ mapped to the largest label $n$ by $\pi$ (i.e., $v = \pi^{-1}(n)$).
    \item Permanently assign $v$ the label $n$ in the hierarchical labeling.
    \item Remove $v$ and recursively apply the procedure to the subgraph $G \setminus \{v\}$ to determine the labels $\{1, \dots, n-1\}$.
\end{enumerate}

By recursively anchoring the largest label according to the base canonizer, we create a new labeling that is guaranteed to be hierarchical by construction, since the canonical form of the $n$-vertex graph is defined by extending the canonical form of its $(n-1)$-vertex subgraph.

\subsection{Integration with SAT Solving (CDCL)}

We integrate this canonicity check directly into the Conflict-Driven Clause Learning (CDCL) loop of the CaDiCaL~\cite{biere2024cadical} SAT solver via a custom propagator~\cite{fazekas2023ipasir}. The problem is encoded using Boolean variables $e_{i,j}$ representing the edges.

\paragraph{The Propagator Logic.}
The SAT solver explores the search space by assigning truth values to edge variables. Although the solver may assign variables in an arbitrary order, our propagator monitors the assignment to detect when a \emph{complete prefix} is formed.
A prefix $G[k]$ is considered complete when all $\binom{k}{2}$ edge variables $e_{i,j}$ with $1 \le i < j \le k$ are assigned.
Whenever the solver completes a new prefix $G[k]$, the propagator invokes the RCL check:
\begin{itemize}
    \item If $G[k]$ is canonical, the search continues.
    \item If $G[k]$ is \textbf{non-canonical}, the propagator adds a blocking clause back to the solver.
\end{itemize}

\paragraph{Blocking Non-Canonical Partials.}
When a prefix $G[k]$ is found to be non-canonical, we generate a \emph{blocking clause} to prune the search branch. Because RCL is hierarchical, the non-canonicity of $G[k]$ implies that no extension to a larger graph $G[n]$ can be canonical.

The blocking clause $C_{\text{block}}$ forbids the specific assignment of edges on the subgraph induced by vertices $\{1, \dots, k\}$. Let $\pi$ be the current partial assignment. For each pair of vertices $1 \le i < j \le k$, we define the literal $\lambda_{i,j}$ representing the negation of the current value of edge variable $e_{i,j}$:
\[
\lambda_{i,j} = 
\begin{cases} 
\neg e_{i,j} & \text{if } e_{i,j} \text{ is assigned \textsc{True} in } \pi, \\
e_{i,j} & \text{if } e_{i,j} \text{ is assigned \textsc{False} in } \pi.
\end{cases}
\]
The blocking clause is then the disjunction of these negated literals:
\[ C_{\text{block}} = \bigvee_{1 \le i < j \le k} \lambda_{i,j}. \]
This clause is added to the solver's database, forcing it to backtrack and modify at least one edge decision within the subgraph $G[k]$.

\section{Orthogonality Check}
\label{sec:orthogonality}

Beyond isomorph-rejection, our search incorporates a custom propagator that exploits the specific geometric nature of the Kochen-Specker problem. Our problem domain provides partial geometric information—specifically, the coordinates of the base set—before the search begins. We leverage this domain knowledge to detect and prune combinatorially valid graphs that are geometrically unrealizable in $\mathbb{C}^3$.

\subsection{Deriving Vector Coordinates}

The search operates by extending a \emph{fixed} subgraph (such as the 25-ray complete SI-C set) to a larger target order (e.g., 33 rays). This creates a distinction between two types of vectors in the system:
\begin{enumerate}
    \item \textbf{Fixed Vectors (Static):} The vectors corresponding to the base subgraph are pre-calculated and fixed. Their coordinates satisfy all orthogonality constraints defined by the subgraph's edges.
    \item \textbf{Search Vectors (Dynamic):} For the remaining vertices (e.g., rays $26$ to $33$), the coordinates are unknown at the start of the search. They are determined dynamically by the SAT solver's choices.
\end{enumerate}

As the SAT solver assigns truth values to edge variables connecting a new vertex $v$ to existing vertices $u_1$ and $u_2$, the propagator infers the geometric consequence. If $v$ is asserted to be orthogonal to both $u_1$ and $u_2$ (where $\vec{u}_1, \vec{u}_2$ are already known), then $\vec{v}$ is uniquely determined (up to a scalar factor) by the cross product: $\vec{v} = \vec{u}_1 \times \vec{u}_2$.
This allows the propagator to constructively determine the coordinates of new rays on-the-fly. If a vertex is connected to only one known vector, its coordinates remain undetermined; once a second orthogonal neighbor is assigned, the coordinates become fixed, triggering further checks.

\subsection{Detecting and Blocking Geometric Violations}\label{subsection:violation}
A geometric contradiction arises when dynamically derived vectors fail to satisfy the required orthogonality relations. We identify two conflict types: (1) \textbf{Generating Parallel Vectors:} A newly derived vector cannot be parallel to existing vectors. (2) \textbf{Violating Orthogonality:} If a derived vector $\vec{v}$ is adjacent to $\vec{w}$ but $\langle \vec{v}, \vec{w} \rangle \neq 0$, the edge constraint is violated.

\paragraph{Isolating the Cause via Dependency Chains.}
When a violation is detected, a naive approach would be to block the entire partial assignment. However, this is inefficient as it does not isolate the specific set of decisions that caused the geometric failure. Instead, we compute a \emph{dependency chain} for every derived vector to construct a minimal blocking clause, which only contains vectors that are involved in causing such violation.

Let $D(v)$ be the set of edge variables responsible for determining the coordinates of vector $\vec{v}$. For a fixed vertex, $D(v) = \emptyset$. For a derived vertex $v$ computed from parents $u_1, u_2$, the dependency set is defined recursively as $D(v) = \{e_{v, u_1}, e_{v, u_2}\} \cup D(u_1) \cup D(u_2)$. When a conflict occurs (e.g., $\vec{v} \cdot \vec{w} \neq 0$ despite edge variable $e_{v,w}$ being true), we learn the clause equivalent to the constraint
\[ C_{\text{block}} = \neg \Biggl( e_{v,w} \land \bigwedge_{e \in D(v)\cup D(w)} e \Biggr) . \]
This clause blocks only the specific chain of edge choices that forced the creation of the incompatible vectors, allowing the solver to retain the valid parts of the partial graph.

\subsection{Exact Arithmetic}
To ensure correctness, we perform all geometric computations using exact arithmetic, avoiding floating-point instability. The arithmetic field is automatically selected based on the input: we use $\mathbb{Z}^3$ for real-valued sets (verifying orthogonality via $\vec{a} \cdot \vec{b} = 0$) and the appropriate algebraic field for complex-valued candidates (using the Hermitian inner product $\langle \vec{a}, \vec{b} \rangle = 0$). Since derived vectors are generated exclusively via cross products, the field remains closed under the operations of the base set.

\section{Verification}
\label{sec:verification}

To ensure the correctness of our exhaustive search, we produce independently verifiable proofs. Our pipeline extends the standard DRAT proof format to support two types of domain-specific axioms: canonicity clauses (t-clauses) and orthogonality clauses (o-clauses). External clauses are labeled as trusted during the standard RUP (Reverse Unit Propagation) check and are independently validated by a simple specialized checker.

\subsection{Proof Format and Trust Boundary}

The proof consists of a sequence of clause additions and deletions. Standard learned clauses are verified via RUP, ensuring they are logical consequences of the current formula. If a standard RUP clause blocks the canonical form, then the problem constraints themselves forbid that specific graph.

The correctness of the external clauses relies on one key principle: a t-clause must never block the unique canonical representative of an isomorphism class. The verifier validates this criteria using a trusted core of graph permutation and exact arithmetic routines (implemented in a lightweight C++ verifier), avoiding dependency on the SAT solver or a Computer Algebra System.

\subsection{Canonicity Clause Verification}

For each t-clause $C$ blocking a graph $G$, the verifier must ensure that this blocking step is sound — that is, that at least one graph isomorphic to $G$ remains reachable by the search. The witness $\pi$ provided in the proof exhibits such a survivor: applying $\pi$ to $G$ yields an isomorphic graph $G' = \pi(G)$, and the verifier confirms that the clause encoding $G'$ and all of its prefixes are not among the blocking clauses. Since permutation preserves the isomorphism class, $G' \cong G$; since $G'$ is unblocked, the iso-class is not eliminated. No lexicographic or ordering relation between $G$ and $G'$ is asserted or checked—in practice $\pi$ is supplied by nauty's recursive canonical labeling, so $G'$ is the canonical representative, but the verifier requires only that $G'$ be some unblocked isomorphic copy.

The verification logic relies on a global consistency check against a hash table $\mathcal{S}$, which stores all t-clauses asserted so far. To validate a clause $C$ with witness $\pi$, the verifier performs the following steps:
\begin{enumerate}
    \item \textbf{Apply Permutation:} Construct the adjacency matrix $M'$ of the isomorphic graph $G'$ using the witness $\pi$ (i.e., $M' = \pi(A_G)$).
    \item \textbf{Check Recursive Validity:} Verify that neither the graph defined by $M'$, nor any of its generating ancestors (prefixes), appears in the blocked set $\mathcal{S}$. 
\end{enumerate}

This check ensures that while $G$ is pruned, its canonical alternative $G'$ (and the path leading to it) remains valid for exploration. Because the maximum graph order is $N=33$ and the search starts from a base of size $p \approx 13$, checking the ancestors of $M'$ requires fewer than $20$ lookups in $\mathcal{S}$ per clause. This limited recursion depth ensures that the verification overhead remains low relative to the solver's search time.

\subsection{Orthogonality Clause Verification}

For o-clauses, the verifier must confirm that the blocked subgraph contains a geometric contradiction. The witness includes the minimal set of edges $E_\text{wit} \subseteq E$ required to derive the contradiction.

The verification proceeds in two steps:
\begin{enumerate}
    \item \textbf{Re-derivation:} The verifier does not trust the vector coordinates provided in the witness. Instead, starting from the fixed base vectors, it \emph{re-computes} the coordinates for the vertices in the witness using the edges in $E_\text{wit}$ and the cross-product rules. This ensures the vectors are strict algebraic consequences of the graph structure.
    \item \textbf{Contradiction Check:} Using the re-derived vectors, the verifier checks for one of two fundamental violations mentioned in Section~\ref{subsection:violation}.
\end{enumerate}

This pipeline ensures that: (1) t-clauses never block at least one representative of every graph arising during the enumeration process (soundness of isomorph rejection), and (2) o-clauses never block a geometrically valid graph (soundness of pruning). Combined with the standard RUP check, this provides a complete certificate that the search space was exhaustively explored.

\section{Results}\label{sec:result}

We present our results in two parts: a performance validation of the \textbf{SAT+\nauty} framework followed by the exhaustive enumeration of Kochen-Specker sets. We begin with a comparison study demonstrating that standard lexicographical symmetry breaking is intractable on the canonical partial assignments dominating this search space. A direct end-to-end comparison is effectively impossible: as noted in Section~\ref{sec:encoding} (Fixed Subgraphs), the lexicographical approach cannot practically canonize even the starting base graph required to initialize the search. The quantitative comparison is reported in Table~\ref{tab:comparison} and visualized in Figures~\ref{fig:comparison-avg}.

\begin{table*}[t]
    \centering
    \caption{Runtime comparison on $100$ lex-least canonical $4$-chromatic $K_4$-free graphs per order.}
    \label{tab:comparison}
    \small
    \setlength{\tabcolsep}{5pt}
    \begin{tabular}{@{}rrrrrrr@{}}
    \toprule
    Order & $N$ & Lex Avg & RCL Avg & Lex Total & RCL Total & Speedup \\
    \midrule
    15 & 100 & 31.9 ms & 26.6 ms & 3.19 s & 2.66 s & 1$\times$ \\
    16 & 100 & 34.6 ms & 26.9 ms & 3.46 s & 2.69 s & 1$\times$ \\
    17 & 100 & 56.5 ms & 27.0 ms & 5.65 s & 2.70 s & 2$\times$ \\
    18 & 100 & 82.4 ms & 26.7 ms & 8.24 s & 2.67 s & 3$\times$ \\
    19 & 100 & 304.2 ms & 26.5 ms & 30.42 s & 2.65 s & 11$\times$ \\
    20 & 100 & 542.7 ms & 27.1 ms & 54.27 s & 2.71 s & 20$\times$ \\
    21 & 100 & 2.51 s & 36.7 ms & 4.2 min & 3.67 s & 68$\times$ \\
    22 & 100 & 6.75 s & 33.7 ms & 11.3 min & 3.37 s & 200$\times$ \\
    23 & 100 & 46.74 s & 29.1 ms & 1.3 hr & 2.91 s & 1{,}606$\times$ \\
    24 & 100 & 51.22 s & 27.9 ms & 1.4 hr & 2.79 s & 1{,}836$\times$ \\
    25 & 100 & 3.2 min & 27.0 ms & 5.3 hr & 2.70 s & 7{,}094$\times$ \\
    26 & 100 & 3.8 min & 26.9 ms & 6.3 hr & 2.69 s & 8{,}499$\times$ \\
    \bottomrule
    \end{tabular}
\end{table*}

\subsection{Performance Analysis: RCL vs. Lex-based}
\label{sec:comparison}

We next isolate the core technical bottleneck addressed by our approach: checking canonicity on the highly structured graphs arising in Kochen-Specker (KS) generation.
As discussed in Section~\ref{sec:encoding}, lexicographical symmetry breaking is attractive because it yields a hereditary (prefix-preserving) notion of canonicity, which enables pruning of partial assignments.
However, in KS-style searches the solver spends most of its time near canonical (and near-canonical) partial graphs, precisely where lex-least checks exhibit their worst-case behavior.
In contrast, our Recursive Canonical Labeling (RCL) construction provides a hereditary canonical form while inheriting the practical efficiency of {\nauty} as its base labeling routine.

\paragraph{Dataset construction: canonical 4-chromatic $K_4$-free graphs.}
To obtain an apples-to-apples comparison on graphs that mimic the structure of KS orthogonality graphs, we generated a benchmark set of graphs constrained to satisfy
(i) connectedness, (ii) the presence of a triangle, (iii) $K_4$-freeness (clique number $\omega(G)=3$), (iv) not 3-colorable ($\chi(G)>3$), and (v) 4-colorable ($\chi(G)\leq4$). While these benchmark graphs are not KS orthogonality graphs, they share similar constraints and structural features (e.g., triangle-rich, $K_4$-free, and highly structured near-canonical instances), making them a useful proxy for stressing canonicity checks in KS-style generation.
For each order $n\in\{15,\ldots,26\}$ we produced $100$ such graphs (for a total of $1{,}200$ instances).

We then transformed each instance into a \emph{lex-least canonical} graph using the standard witness-guided procedure: repeatedly run the lex-least canonicity checker; if it returns a witness permutation exhibiting an isomorphic but lexicographically smaller labeling, relabel the graph accordingly and iterate until no improving witness exists.
In practice, this canonization process becomes prohibitively expensive beyond this range; we stop at $n=26$ because many graphs of order $27$ required hours (or timed out) to reach the lex-least canonical representative.

\begin{figure}[!t]
    \centering
    \includegraphics[width=\linewidth]{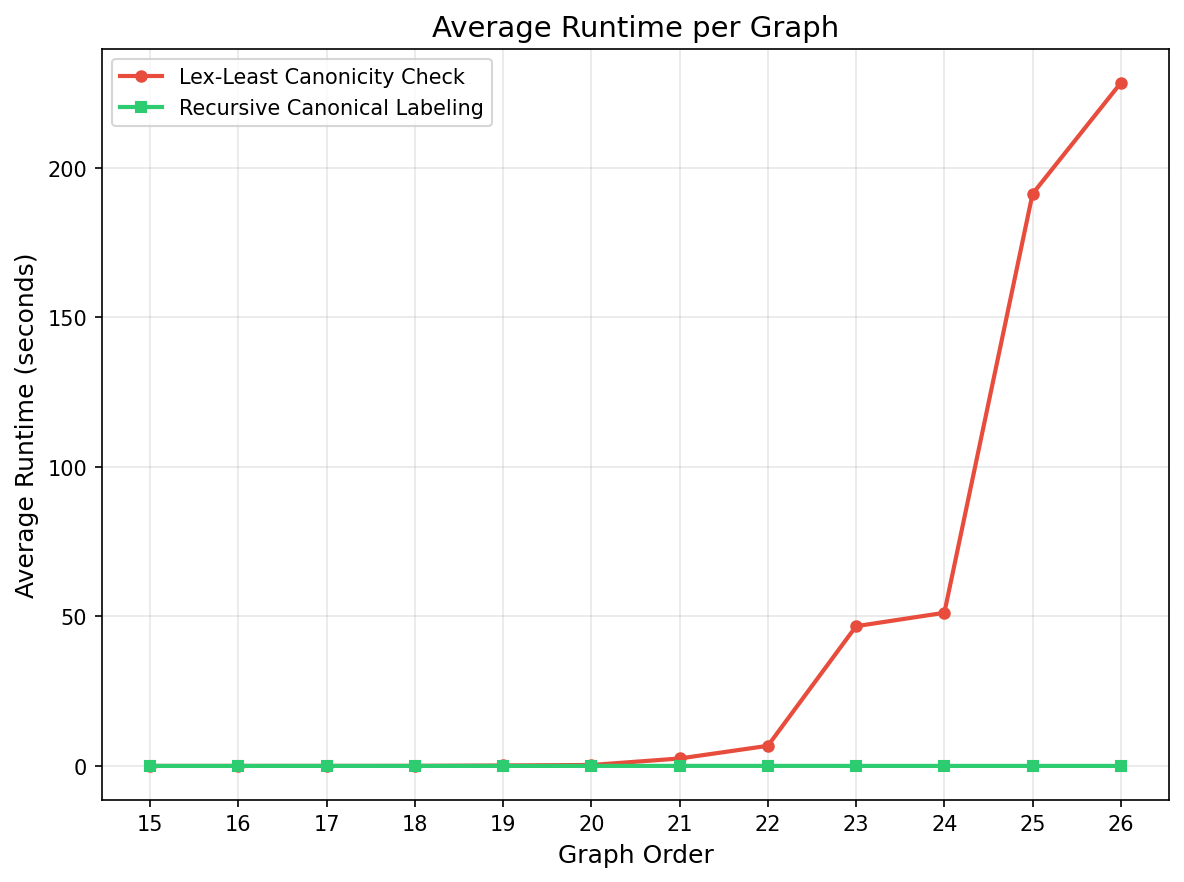}
    \caption{Average canonicity-check runtime on the benchmark dataset (same data as Table~\ref{tab:comparison}).}
    \label{fig:comparison-avg}
\end{figure}

\paragraph{Experimental comparison.}
On this dataset of lex-least canonical graphs, we measured the runtime of (a) the lex-least canonicity check itself and (b) our RCL canonicity check (with {\nauty} as the base labeling).
Figure~\ref{fig:comparison-avg} summarizes the results.
The lex-least check scales exponentially with $n$: its mean runtime increases from roughly $32$ms at $n=15$ to about $3.8$ minutes at $n=26$ (approximately a $7{,}000\times$ increase).
By contrast, RCL remains essentially flat at around $27$ms throughout, i.e., effectively independent of graph size in this range.

The performance gap accelerates rapidly after $n\approx 21$.
At $n=20$ the speedup is a modest $\sim 20\times$, but it jumps to $\sim 200\times$ at $n=22$, $1{,}606\times$ at $n=23$, and reaches $8{,}499\times$ at $n=26$.
Moreover, RCL's total time across all $100$ graphs at a fixed order (about $2.7$s) is less than what the lex-least check spends on a \emph{single} canonical graph at $n=26$ (about $3.8$ minutes).
These measurements explain why lexicographical symmetry breaking becomes the dominant bottleneck in our target instances, and why replacing the lex-least checker with RCL is necessary to make the exhaustive enumeration in Section~\ref{sec:results-recursive} feasible.

\subsection{Exhaustive Enumeration of KS Sets Extending the Complete SI-C Set}
\label{sec:results-recursive}

Using our RCL-based solver, we performed an exhaustive search for Kochen-Specker sets extending the $25$-ray complete SI-C set. The search space was partitioned into $128$ independent sub-problems using AlphaMapleSAT~\cite{AlphaMapleSAT:2024} and executed on a high-performance cluster of Dual AMD EPYC 7713 processors.

The enumeration for order $33$ consumed $1{,}641$ hours of CPU time (approximately $68$ days), which was completed in roughly $2$ days of wall-clock time across $128$ cores. The solver identified $44$ non-isomorphic candidates that satisfy the combinatorial Kochen-Specker conditions. However, the embeddability check post-processing step confirmed that only \textbf{one} of these candidates admits an orthogonal representation in either $\mathbb{R}^3$ or $\mathbb{C}^3$. This unique solution is isomorphic to the $33$-ray set discovered by Schütte \cite{Bub:1996FP}; thus, we formally confirm that, up to order $33$, the Schütte set is the \emph{only} Kochen-Specker set containing the minimal SI-C set of order $25$. 

To ensure the correctness of this uniqueness result, the entire search process was formally certified. The solver produced a comprehensive proof trace in an extended DRAT format, recording all learned clauses alongside domain-specific axioms for isomorph-rejection (t-clauses) and geometric pruning (o-clauses). For the order-$33$ enumeration, the total proof size was approximately $13$ TiB\@. This certificate was independently validated by a custom verifier implemented as a standalone Python script. While the verifier code is not formally proven, it is designed to be minimal and auditable, reducing the trusted computing base to a simple set of logical checks that provide a rigorous guarantee that no valid Kochen-Specker sets were overlooked.

\section{Conclusion}

We presented a SAT-based orderly generation framework that integrates {\nauty} to overcome the scalability limitations of lexicographical symmetry breaking.
Our analysis confirms that while standard lexicographical checks suffice for random graphs, they incur exponential overhead on the highly structured subgraphs inherent to combinatorial search problems, resulting in a performance gap of nearly four orders of magnitude compared to RCL\@.
Because the search for a canonical solution inevitably requires the verification of canonical and near-canonical partial graphs, standard lexicographical checkers are forced into their worst-case performance.
By adopting Recursive Canonical Labeling (RCL), we achieved negligible isomorphism overhead, enabling the first exhaustive enumeration of Kochen-Specker sets extending the $25$-ray SI-C closure up to order $33$.
This framework allowed us to settle the uniqueness of the Sch\"utte set relative to the complete $25$-ray core. The SAT+{\nauty} framework offers a practical solution for combinatorial generation problems where symmetry handling is the primary bottleneck and can generalize to other domains requiring isomorph-free generation of complex combinatorial structures.
\bibliographystyle{named}
\bibliography{ijcai26}

\begin{thebibliography}{}

\bibitem[\protect\citeauthoryear{Afzaly}{2016}]{afzaly2016generation}
Seyedeh~Narjess Afzaly.
\newblock {\em Generation of graph classes with efficient isomorph rejection}.
\newblock PhD thesis, Australian National University, 2016.

\bibitem[\protect\citeauthoryear{Aolita \bgroup \em et al.\egroup
  }{2012}]{Aolita:2012PRA}
Leandro Aolita, Rodrigo Gallego, Antonio Ac{\'\i}n, Andrea Chiuri, Giuseppe
  Vallone, Paolo Mataloni, and Ad{\'a}n Cabello.
\newblock Fully nonlocal quantum correlations.
\newblock {\em Phys. Rev. A}, 85:032107, 2012.

\bibitem[\protect\citeauthoryear{Aravind \bgroup \em et al.\egroup
  }{2025}]{Aravind:2025}
Padmanabhan~K Aravind, Justin~YJ Burton, David Richter, and Guillermo
  Nu{\~n}ez~Ponasso.
\newblock Triacontagonal proofs of the {Bell-Kochen-Specker} theorem.
\newblock {\em Journal of Physics A: Mathematical and Theoretical}, 2025.

\bibitem[\protect\citeauthoryear{Bengtsson \bgroup \em et al.\egroup
  }{2012}]{Bengtsson:2012PLA}
I.~Bengtsson, K.~Blanchfield, and A.~Cabello.
\newblock A {K}ochen--{S}pecker inequality from a {SIC}.
\newblock {\em Phys. Lett. A}, 376:374--376, 2012.

\bibitem[\protect\citeauthoryear{Biere \bgroup \em et al.\egroup
  }{2024}]{biere2024cadical}
Armin Biere, Tobias Faller, Katalin Fazekas, Mathias Fleury, Nils Froleyks, and
  Florian Pollitt.
\newblock Cadical 2.0.
\newblock In {\em International Conference on Computer Aided Verification},
  pages 133--152. Springer, 2024.

\bibitem[\protect\citeauthoryear{Bright \bgroup \em et al.\egroup
  }{2021}]{Bright2021}
Curtis Bright, Kevin K.~H. Cheung, Brett Stevens, Ilias Kotsireas, and Vijay
  Ganesh.
\newblock A {SAT}-based resolution of {Lam's} problem.
\newblock {\em Proceedings of the AAAI Conference on Artificial Intelligence},
  35(5):3669–3676, May 2021.

\bibitem[\protect\citeauthoryear{Bright \bgroup \em et al.\egroup
  }{2022}]{bright2022satisfiability}
Curtis Bright, Ilias Kotsireas, and Vijay Ganesh.
\newblock When satisfiability solving meets symbolic computation.
\newblock {\em Communications of the ACM}, 65(7):64--72, 2022.

\bibitem[\protect\citeauthoryear{Bub}{1996}]{Bub:1996FP}
Jeffrey Bub.
\newblock Sch\"{u}tte's tautology and the {K}ochen-{S}pecker theorem.
\newblock {\em Found. Phys.}, 26:787--806, 1996.

\bibitem[\protect\citeauthoryear{Cabello \bgroup \em et al.\egroup
  }{2016}]{Cabello:2016JPA}
A.~Cabello, M.~Kleinmann, and J.~R. Portillo.
\newblock Quantum state-independent contextuality requires 13 rays.
\newblock {\em J. Phys. A: Math. Theor.}, 49:38LT01, 2016.

\bibitem[\protect\citeauthoryear{Cabello}{1996}]{Cabello:1996}
Ad\'an Cabello.
\newblock {\em Pruebas algebraicas de imposibilidad de variables ocultas en
  mec\'anica cu\'antica}.
\newblock PhD thesis, Universidad Complutense de Madrid, 1996.

\bibitem[\protect\citeauthoryear{Cabello}{2025a}]{Cabello:2025PRL}
Ad\'an Cabello.
\newblock Simplest bipartite perfect quantum strategies.
\newblock {\em Phys. Rev. Lett.}, 134:010201, Jan 2025.

\bibitem[\protect\citeauthoryear{Cabello}{2025b}]{Cabello:2025XXX}
Adán Cabello.
\newblock The simplest {K}ochen-{S}pecker set.
\newblock {\em Phys. Rev. Lett.}, 135:190203, 2025.

\bibitem[\protect\citeauthoryear{Cinelli \bgroup \em et al.\egroup
  }{2005}]{CinelliPRL2005}
C.~Cinelli, M.~Barbieri, R.~Perris, P.~Mataloni, and F.~De~Martini.
\newblock {A}ll-{V}ersus-{N}othing {N}onlocality {T}est of {Q}uantum
  {M}echanics by {T}wo-{P}hoton {H}yperentanglement.
\newblock {\em Phys. Rev. Lett.}, 95:240405, 2005.

\bibitem[\protect\citeauthoryear{Crawford \bgroup \em et al.\egroup
  }{1996}]{crawford1996symmetry}
James~M. Crawford, Matthew~L. Ginsberg, Eugene~M. Luks, and Amitabha Roy.
\newblock Symmetry-breaking predicates for search problems.
\newblock In {\em Proceedings of the Fifth International Conference on
  Principles of Knowledge Representation and Reasoning}, KR'96, page 148–159,
  San Francisco, CA, USA, 1996. Morgan Kaufmann Publishers Inc.

\bibitem[\protect\citeauthoryear{Farad{\v{z}}ev}{1978}]{faradvzev1978constructive}
I~A Farad{\v{z}}ev.
\newblock Constructive enumeration of combinatorial objects.
\newblock In {\em Probl{\`e}mes combinatoires et th{\'e}orie des graphes},
  pages 131--135, 1978.

\bibitem[\protect\citeauthoryear{Fazekas \bgroup \em et al.\egroup
  }{2023}]{fazekas2023ipasir}
Katalin Fazekas, Aina Niemetz, Mathias Preiner, Markus Kirchweger, Stefan
  Szeider, and Armin Biere.
\newblock {IPASIR-UP}: User propagators for {CDCL}.
\newblock In Meena Mahajan and Friedrich Slivovsky, editors, {\em 26th
  International Conference on Theory and Applications of Satisfiability Testing
  (SAT 2023)}, volume 271 of {\em Leibniz International Proceedings in
  Informatics (LIPIcs)}, pages 8:1--8:13. Schloss Dagstuhl -- Leibniz-Zentrum
  f\"{u}r Informatik, 2023.

\bibitem[\protect\citeauthoryear{Heule}{2019}]{Heule2019}
Marijn J.~H. Heule.
\newblock Optimal symmetry breaking for graph problems.
\newblock {\em Mathematics in Computer Science}, 13(4):533--548, May 2019.

\bibitem[\protect\citeauthoryear{Jha \bgroup \em et al.\egroup
  }{2024}]{AlphaMapleSAT:2024}
Piyush Jha, Zhengyu Li, Zhengyang Lu, Curtis Bright, and Vijay Ganesh.
\newblock {AlphaMapleSAT}: An {MCTS}-based cube-and-conquer {SAT} solver for
  hard combinatorial problems, 2024.

\bibitem[\protect\citeauthoryear{Kernaghan}{2026}]{Kernaghan:2026}
Michael Kernaghan.
\newblock The algebraic landscape of {Kochen-Specker} sets in dimension three.
\newblock {\em arXiv preprint arXiv:2603.16988}, 2026.

\bibitem[\protect\citeauthoryear{Kirchweger and Szeider}{2024}]{KirchwegerS24}
Markus Kirchweger and Stefan Szeider.
\newblock {SAT Modulo Symmetries} for graph generation and enumeration.
\newblock {\em {ACM} Trans. Comput. Log.}, 25(3), 2024.

\bibitem[\protect\citeauthoryear{Kirchweger \bgroup \em et al.\egroup
  }{2023}]{Kirchweger:2023}
Markus Kirchweger, Tomáš Peitl, and Stefan Szeider.
\newblock Co-certificate learning with {SAT} modulo symmetries.
\newblock In Edith Elkind, editor, {\em Proceedings of the Thirty-Second
  International Joint Conference on Artificial Intelligence, {IJCAI-23}}, pages
  1944--1953. International Joint Conferences on Artificial Intelligence
  Organization, 2023.

\bibitem[\protect\citeauthoryear{Kochen and Specker}{1965}]{Kochen:1965a}
S.~Kochen and E.~P. Specker.
\newblock Logical structures arising in quantum theory.
\newblock In J.~W. Addison, L.~Henkin, and A.~Tarski, editors, {\em Symposium
  on the Theory of Models: Proceedings of the 1963 International Symposium at
  Berkeley}, pages 177--189. North-Holland, Amsterdam, 1965.

\bibitem[\protect\citeauthoryear{Kochen and Specker}{1967}]{Kochen:1967JMM}
Simon Kochen and Ernst~P. Specker.
\newblock The {P}roblem of {H}idden {V}ariables in {Q}uantum {M}echanics.
\newblock {\em J. Math. Mech.}, 17:59--87, 1967.

\bibitem[\protect\citeauthoryear{Kumar \bgroup \em et al.\egroup
  }{2025}]{kumar2025}
Shashwat Kumar, Eliott Rosenberg, et~al.
\newblock Quantum-classical separation in bounded-resource tasks arising from
  measurement contextuality, 2025.
\newblock arXiv:2512.02284.

\bibitem[\protect\citeauthoryear{Li \bgroup \em et al.\egroup }{2024}]{Li:2024}
Zhengyu Li, Curtis Bright, and Vijay Ganesh.
\newblock A {SAT} solver + computer algebra attack on the minimum
  {K}ochen–{S}pecker problem.
\newblock In Kate Larson, editor, {\em Proceedings of the Thirty-Third
  International Joint Conference on Artificial Intelligence, {IJCAI-24}}, pages
  1898--1906. International Joint Conferences on Artificial Intelligence
  Organization, 2024.

\bibitem[\protect\citeauthoryear{Liang \bgroup \em et al.\egroup
  }{2016}]{liang2016learning}
Jia~Hui Liang, Vijay Ganesh, Pascal Poupart, and Krzysztof Czarnecki.
\newblock Learning rate based branching heuristic for {SAT} solvers.
\newblock In {\em International Conference on Theory and Applications of
  Satisfiability Testing}, pages 123--140. Springer, 2016.

\bibitem[\protect\citeauthoryear{McKay and Piperno}{2014}]{MCKAY201494}
Brendan~D. McKay and Adolfo Piperno.
\newblock Practical graph isomorphism, {II}.
\newblock {\em Journal of Symbolic Computation}, 60:94--112, 2014.

\bibitem[\protect\citeauthoryear{McKay}{2007}]{mckay2007nauty}
Brendan~D McKay.
\newblock Nauty user’s guide (version 2.4).
\newblock {\em Computer Science Dept., Australian National University}, pages
  225--239, 2007.

\bibitem[\protect\citeauthoryear{Meringer}{1999}]{meringer1999fast}
Markus Meringer.
\newblock Fast generation of regular graphs and construction of cages.
\newblock {\em Journal of Graph Theory}, 30(2):137--146, 1999.

\bibitem[\protect\citeauthoryear{Metin \bgroup \em et al.\egroup
  }{2018}]{Metin2018}
Hakan Metin, Souheib Baarir, Maximilien Colange, and Fabrice Kordon.
\newblock {CDCLSym}: Introducing effective symmetry breaking in {SAT} solving.
\newblock In {\em Tools and Algorithms for the Construction and Analysis of
  Systems}, pages 99--114, New York, 2018. Springer International Publishing.

\bibitem[\protect\citeauthoryear{Peres}{1993}]{Peres:1993}
Asher Peres.
\newblock {\em Quantum Theory: Concepts and Methods}.
\newblock Kluwer, Dordrecht, 1993.

\bibitem[\protect\citeauthoryear{Read}{1978}]{read1978every}
Ronald~C Read.
\newblock Every one a winner or how to avoid isomorphism search when
  cataloguing combinatorial configurations.
\newblock {\em Annals of Discrete Mathematics}, 2:107--120, 1978.

\bibitem[\protect\citeauthoryear{Sellmann and
  Hentenryck}{2005}]{DBLP:conf/ijcai/SellmannH05}
Meinolf Sellmann and Pascal~Van Hentenryck.
\newblock Structural symmetry breaking.
\newblock In Leslie~Pack Kaelbling and Alessandro Saffiotti, editors, {\em
  IJCAI-05}, pages 298--303, California, 2005.

\bibitem[\protect\citeauthoryear{Trandafir and
  Cabello}{2025}]{Trandafir:2025PRAa}
Stefan Trandafir and Ad\'an Cabello.
\newblock Two fundamental solutions to the rigid {K}ochen-{S}pecker set problem
  and the solution to the minimal {K}ochen-{S}pecker set problem under one
  assumption.
\newblock {\em Phys. Rev. A}, 111:052204, May 2025.

\bibitem[\protect\citeauthoryear{Xu \bgroup \em et al.\egroup
  }{2022}]{Xu:2022PRL}
Jia-Min Xu, Yi-Zheng Zhen, Yu-Xiang Yang, Zi-Mo Cheng, Zhi-Cheng Ren, Kai Chen,
  Xi-Lin Wang, and Hui-Tian Wang.
\newblock Experimental {D}emonstration of {Q}uantum {P}seudotelepathy.
\newblock {\em Phys. Rev. Lett.}, 129:050402, Jul 2022.

\bibitem[\protect\citeauthoryear{Xu \bgroup \em et al.\egroup
  }{2024}]{Xu:2024PRL}
Zhen-Peng Xu, Debashis Saha, Kishor Bharti, and Ad\'an Cabello.
\newblock Certifying sets of quantum observables with any full-rank state.
\newblock {\em Phys. Rev. Lett.}, 132:140201, Apr 2024.

\bibitem[\protect\citeauthoryear{Yang \bgroup \em et al.\egroup
  }{2005}]{YangPRL2005}
Tao Yang, Qiang Zhang, Jun Zhang, Juan Yin, Zhi Zhao, Marek {\.Z}ukowski,
  Zeng-Bing Chen, and Jian-Wei Pan.
\newblock {A}ll-{V}ersus-{N}othing {V}iolation of {L}ocal {R}ealism by
  {T}wo-{P}hoton, {F}our-{D}imensional {E}ntanglement.
\newblock {\em Phys. Rev. Lett.}, 95:240406, 2005.

\bibitem[\protect\citeauthoryear{Yu and Oh}{2012}]{Yu:2012PRL}
Sixia Yu and C.~H. Oh.
\newblock State-independent proof of {K}ochen-{S}pecker theorem with 13 rays.
\newblock {\em Phys. Rev. Lett.}, 108:030402, 2012.

\end{thebibliography}

\end{document}